\documentclass[preprint,showpacs,preprintnumbers,amsmath,amssymb]{revtex4}
\usepackage{graphicx,color}

\begin{document}
\title{Laser-induced instabilities in liquid crystal cells with a photosensitive substrate}
\author{Istv\'an J\'anossy\footnote{Corresponding author;
janossy.istvan@wigner.mta.hu}, Katalin Fodor-Csorba, Anik\'o
Vajda, and Tibor T\'oth-Katona} \affiliation{Institute for Solid
State Physics and Optics, Wigner Research Centre for Physics,
Hungarian Academy of Sciences, H-1525 Budapest, P.O.Box 49,
Hungary}

\date{\today}

\begin{abstract}
Liquid crystal layers sandwiched between a reference plate and a
photosensitive substrate were investigated. We focused on the
reverse geometry, where the cell was illuminated by a laser beam
from the reference side. In planar cells instabilities occurred,
static and dynamic ones, depending on the angle between the laser
polarization and the director orientation on the reference plate.
In cells where the molecules were aligned along the normal of the
reference plate, dynamic pattern was observed at all angles of
polarization. A simple model based on a photo-induced surface
torque gives account of the findings. Light scattering studies
revealed some basic properties of the instabilities.
\end{abstract}

\pacs{61.30.-v, 61.30.Hn, 78.15.+e, 42.70.Df} \maketitle

\section{Introduction}

Optical reorientation of liquid crystals is known since decades
\cite{Mar}. In a typical experiment the initial director alignment
in an oriented cell transforms into a new steady-state
configuration under the influence of a light beam. There are,
however, some cases when steady-state director pattern does not
exist or it is not stable, even though the exciting beam is
uniform both in time and space. Temporal as well as spatial
instabilities were observed. As an example, temporal oscillations
and spontaneous spatial pattern formation were found in
Fabry-Perot resonators containing nematic liquid crystals
\cite{Cheung,Kreuzer}. Irradiation of homeotropic films of liquid
crystals with circularly polarized beam is known to cause undamped
director precession \cite{San,Zol1,Brass}; in the same cells
ordinarily polarized light beam with slightly oblique incidence
induces chaotic director oscillations \cite{Zol2,Cip,Brass1}.

In the above-mentioned examples the source of director
reorientation is the bulk torque, exerted by the electromagnetic
field on the liquid crystal molecules. There is another mechanism
of optical reorientation, namely photo-induced surface realignment
of the director \cite{Gib,Ich}. This effect occurs when one of the
plates of the liquid crystal cell is coated with a photosensitive
material. Most often azo dyes are used as a coating substance,
exhibiting \emph{trans-cis} photoisomerization. When the cell is
irradiated, the director on the photosensitive plate becomes
oriented perpendicularly to the light polarization at the plate.

In a recent publication \cite{Jan} we investigated cells, in which
the liquid crystal was sandwiched between a "reference" plate,
covered with a traditional, rubbed polyimide layer and a
dye-coated plate. We found that instabilities occur in the liquid
crystal film when the light beam enters the cell from the
reference side (\emph{reverse} geometry). In the experiment the
polarized light from a microscope was used as the light source for
excitation. The dominant instability was a static spatial pattern
formation with a characteristic domain size of the order of a
micrometer. At certain locations of the sample, however, dynamic
instability was observed. In this structure the director
oscillated or rotated continuously in a spatially inhomogeneous
manner. The oscillation period was found to be proportional to the
intensity of the illuminating light.

In the present communication further results are described in
connection with this instability. We report on experiments where a
monochromatic laser beam was applied for excitation instead of the
white-light source used previously. In planar cells at ordinary
input of the exciting light (polarization perpendicular to the
rubbing direction at the reference plate) the director
configuration remained stable. At extraordinary input
(polarization parallel to rubbing) static light-scattering
occurred with a faint and broad conical maximum around a certain
angle. When the angle between the light polarization and the
rubbing direction was between $o$ and $e$ polarization, dynamic
scattering developed. We present also results obtained in a cell
where on the reference plate the liquid crystal molecules were
aligned perpendicularly to the surface (so-called hybrid cell). In
this type of cell dynamic behavior was observed at all angles of
incident polarization. These experimental results are described in
detail in Section II, while in Section III a simple model is
presented, which provides an explanation of the findings. In
addition, we studied the fluctuations of the scattered light
(Section IV). In planar cells intermittency, i.e., the occurrence
of relatively rare fluctuations of huge amplitudes were observed.
The probability density function (PDF) of the scattered light
intensity could be characterized by the generalized
Fisher-Tippet-Gumbel distribution (gFTG) \cite{Bram}. In the case
of a hybrid cell neither gFTG nor Gaussian distributions did not
provide a good description of the fluctuations; the PDF was
between these two distribution functions.

\section{Basic observations in the reverse geometry}

The photosensitive plate was prepared in the same way as in our
previous work \cite{Jan} following the method worked out by Yi et
al. \cite{Yi}. The azo-dye utilized in the experiments was
synthesized by coupling to methyl-red (3-aminopropyl)
triethoxysilane, forming a derivatized methyl red (dMR). The dMR
was chemisorbed on the glass surface in toluene solution. To
prepare planar cells commercial, rubbed polyimide coated slides
from E.H.C. Co (Japan) were applied as reference plates. $15-20
\mu$m thick cells have been filled with 4-cyano-4'-pentylbiphenyl
(5CB), which is a room temperature nematic. Before and during the
filling process the cell was illuminated with polarized light from
a white LED source from the photosensitive side; the polarization
direction was perpendicular to the rubbing direction. This
procedure ensured good quality planar alignment of the cell. The
preparation process of hybrid cells was similar to that of the
planar one, except that the reference plate was coated with
L-a-phosphatidylcholine (lecithin), which ensured homeotropic
director orientation on that plate. Prior to and during filling
the cell with the liquid crystal it was irradiated again with
polarized light. As a result, the initial director field was
confined to the plane defined by the cell normal and the
polarization direction of the irradiating light. The measurements
were carried out at room temperature. The laser beam was focused
to around $100\mu$m spot size; its power was changed from a few
$\mu$W to a few mW.

For excitation the beams from a green ($\lambda = 532$nm) and a
blue ($\lambda = 457$nm) laser were used. The results were similar
for the two wavelengths; in the paper data obtained with the green
laser are presented. In the geometry, when the dye-coated plate
was facing the illuminating beam (\emph{direct} geometry), a
simple reorientation process was observed. In this case the
direction of the light polarization on the photosensitive
substrate obviously coincides with the impinging laser
polarization (on the contrary this is not the case for the reverse
geometry.  The director rotates on the dyed substrate toward the
normal of the incident polarization, while on the reference plate
it remains unchanged thereby inducing a twist in the cell
\cite{footnote1}.

The subject of the present paper is the investigation of the
\emph{reverse} geometry, i.e., the case when the cell is
illuminated from the reference plate. First we describe the
observations in planar cells; subsequently we discuss findings in
hybrid cells.

In planar cells qualitatively different behaviors were observed
depending on the angle between the rubbing direction and the
polarization of the incoming light. We represent this angle by
$\alpha$, where $\alpha=0$ denotes that the polarization is
perpendicular to the rubbing direction. At $\alpha=0^\circ$ the
beam traversed the sample without any significant depolarization
or scattering. This was as expected, since in this case the
initial planar alignment on the photosensitive plate is stabilized
by the illuminating beam. At $\alpha = 90^\circ$, i.e., when the
incoming polarization was parallel to the director on the
reference plate, static scattering took place. The scattered light
exhibited a broad peak along a cone; for the green beam the
aperture was about 20-30 degrees. This scattering obviously
corresponded to the static domain structure, observed previously
in a microscope \cite{Jan}. From the aperture of the scattering
cone one can estimate the domain size. It was found to be in the
order of few micrometers, in good agreement with the findings in a
microscope. At angles between $0^\circ$ and $90^\circ$ dynamic
scattering arose in the form of random light intensity both in
space and in time. In the scattering picture flashes of randomly
oriented lines were observed as well. It was difficult to assign
exact critical angles to transitions from non-scattering or static
scattering regions to dynamically scattering ones; usually the
interval $15^\circ < \alpha < 75^\circ$ belonged to the latter
type of instability. This mode of instability corresponds to the
dynamical instability, observed at certain locations of the sample
under a polarizing microscope at $90^\circ$ \cite{Jan}. However,
in the present case, using a monochromatic light source, the
turbulent director rotation was observed in the major part of the
sample and at a certain range of the $\alpha$ values.

In order to characterize the dynamic nature of the instability, we
recorded the temporal fluctuations of the scattered light $S$. A
fiber optics was used to collect and to transmit to a
photomultiplier a small part of the scattered light. The fiber was
placed around the center of the halo observed at the static case
(i.e. $15^\circ$ away from the direct beam) and collected the
scattered light from a cone with approximately $2^\circ$ aperture.
The sampling rate was around 5 Hz; the integration time of the
intensity measurements was 20 ms. As an example, a trace observed
at $\alpha=45^\circ$ is presented in Fig. 1. As a comparison, we
show also in Fig. 1 the corresponding curve for $\alpha=90^\circ$,
where no significant fluctuations were found. The detailed
analysis of the scattered light fluctuations is presented in
Section IV.

Observations for the direct geometry in the hybrid cells will be
discussed elsewhere. In the reverse geometry dynamic instability
was found in these cells, both using microscope light and laser
beams. The instability was dynamic at arbitrary $\alpha$  angle,
although it showed some dependence on this angle. The temporal
fluctuations of the scattered light are displayed in Fig. 2 for
three different values. The turbulence was less pronounced in the
hybrid cells compared to that in planar ones: here less intense
spikes were observed much more rarely than in the latter case.

\section{Mechanism of the instabilities}

In the direct geometry the orientation of the liquid crystal
molecules on the photosensitive plate is determined by the
incoming polarization direction of the exciting light beam. In the
reverse geometry the exciting radiation passes through the liquid
crystal layer  before impinging the photosensitive layer. While
traversing the sample the state of polarization of the light wave
can be modified due to the birefringence of the substance; the
modification depends on the director pattern. In turn, the state
of polarization at the photosensitive plate determines the surface
director on this plate and thus the director distribution in the
cell. Therefore, in the reverse geometry the photoalignment and
the director field form a coupled system mutually influencing each
other;  as it will be demonstrated below,  this coupling is the
reason behind the pattern formation.

The exact description of the dynamics of photoalignment and the
corresponding director reorientation is a complex problem, which
in our opinion is not yet fully solved. An approach was proposed
by Kiselev et al \cite{Kis}, who associated photoalignment with a
light intensity-dependent mean-field potential. In the present
model we adopt this approach. In this paper we confine ourselves
to monochromatic exciting light; the case of white light source
will be analyzed elsewhere. We prefer to describe photoalignment
through a light-induced surface torque on the director on the dyed
plate, which can be derived from the mean-field potential, because
it is straightforward to generalize it to elliptically polarized
light. In the simplest form that is compatible with the symmetry
of the problem it can be written in a similar form as the bulk
optical torque, with the difference that it acts only at the
photosensitive boundary:
\begin{equation}\label{torque}
    \mathbf{\Gamma}_{ph}=f<(\mathbf{n}_s\times\mathbf{E}_s)\mathbf{n}_s\cdot\mathbf{E}_s>
\end{equation}
where $\mathbf{E}_s$ and $\mathbf{n}_s$ are the electric field
strength and the surface director on the photosensitive plate
respectively; $f$ is a positive constant and $<\,>$ denotes time
averaging over a period of oscillation of the electromagnetic
wave. The above conjecture provides a qualitatively correct
description of photo-alignment, including elliptically polarized
exciting beams. For elliptic polarization the surface torque can
be written as
\begin{equation}\label{elltorque}
    \mathbf{\Gamma}_{ph}=f E_oE_e\cos \Delta \Phi(\mathbf{n}_s\times\mathbf{e}_o)(\mathbf{n}_s\cdot\mathbf{e}_e)
\end{equation}
where $E_o$ and $E_e$ are the amplitude of the ordinary and
extraordinary components respectively; $\Delta\Phi$ is the phase
difference between the $e$ and $o$ components; $\mathbf{e}_o$ and
$\mathbf{e}_e$ are unit vectors along the ordinary and
extraordinary polarization.

The next assumption is that the light propagates according to the
Mauguin limit \cite{Vries}, i.e., the polarization ellipse follows
adiabatically the rotation of the director within the sample in
the presence of a twist deformation.

First we discuss planar cells. In the case of normal incidence the
surface torque rotates the director within the photosensitive
plate. When the input light on the reference plate is linearly
polarized at an angle $\alpha$ in the planar cell the photoinduced
surface torque given by Eq. (\ref{elltorque}) becomes
\begin{equation}\label{torque2}
      \Gamma_{ph}=f'I\sin 2 \alpha\cos \Delta \Phi
\end{equation}
with $\Delta\Phi=2\pi(n_e-n_o)d/\lambda$. Here $I$ is the
intensity; $n_e$ and $n_o$ are the extraordinary and ordinary
refractive indices, respectively; $d$ and $\lambda$  are the
sample thickness and wavelength respectively; $f'$ is a constant
related to $f$.

Except of special cases, to be discussed below, the photoinduced
torque is different from zero; therefore the director should
rotate either clockwise or anti-clockwise, depending on the sign
of $\cos \Delta\Phi$. Within the Mauguin limit the light
polarization ellipse on the photosensitive plate is rotating
together with the surface director, therefore the photoinduced
torque remains constant in time. This torque is opposed by the
surface elastic torque, which increases as the director rotation
increases the twist in the cell. As shown, however, by experiments
in the direct geometry \cite{Jan}, before a balance would be
reached between the surface torques a disclination loop is formed
in the sample, which reduces the twist. In this way the elastic
torque is reduced, hence balance between torques is never
established. This mechanism explains the observed dynamical
behavior.

At $\alpha =0^{\circ}$ and $\alpha = 90^{\circ}$ the photoinduced
torque is zero. In the former case the director configuration is
stable, while in the latter case it is unstable against small
deviations from the perfect geometry. The second case show a
certain analogy to the electric-field induced Freedericksz
transition in a planar cell, where - in an ideal case - the
transition is initiated by thermal fluctuations of the director.
These fluctuations relieve the twofold degeneracy of the director
rotation and lead to the formation of domains with opposite
twists. The domains are separated by inversion walls, which are
unstable, so finally, the domains merge and a uniform director
rotation takes place. In our case, however, the situation is
different from the electric-field induced Freedericksz transition
in important aspects. First, static disorder of the alignment on
the reference plate plays a much more significant role than
thermal fluctuations as shown e.g. by Nespoulous et al.
\cite{Nesp}. These authors studied SiO substrates and found a few
degrees of spatial disorder in the alignment. The variation of the
orientation on the reference plate results again in the formation
of two types of domains, but this time the domains cannot easily
merge because the walls separating them are attached to the
substrate. In addition, the formation of a fine domain structure
leads to the invalidation of the adiabatic light propagation; the
light beam becomes depolarized on the photosensitive plate, which
weakens its orienting strength, so that the elastic torque can
balance the photoinduced one. These factors can results in the
static spatial structure observed in the experiments.

We note that there is one more situation when the light-induced
torque is zero: the case of $\cos\Delta\Phi=0$, i.e., when an
input beam with a linear polarization at $\alpha=45^\circ$ becomes
circularly polarized at the photosensitive layer. This occurs at
special thicknesses of the sample obeying the relation
\begin{equation}\label{thickness}
 d=(m+\frac{1}{2})\frac{ \lambda}{2(n_e-n_o)}
\end{equation}
where $m$ is an integer. We plan to check in the future such
situations \cite{footnote2}.

In the case of hybrid cells, for the interval $0<\alpha <
90^\circ$, the same argument can be used as for the planar case.
Regarding the situation  $\alpha=0$ and $\alpha = 90^\circ$, we
note that the disorder of orientation on the reference plate is
not present, or at least it is much smaller than in the case of
planar orientation. This may explain the fact that in these cells
no static domains were found. Actually, there is an axial symmetry
of the sample around the $z$ axis, which would suggest that the
scattering is independent from $\alpha$. This argument is valid,
however, only if there is no "easy axis", i.e., preferred director
orientation on the photosensitive plate. Such a preferred axis,
however, might be generated during the preparation of the cell; it
may correspond to the normal direction to the light polarization
used during the cell preparation. This can be one of the reasons
of the observed deviations in the scattering strength for the
three curves presented in Fig. 2.

To sum up, the reason of the dynamic instability is - similarly to
the case of bulk instabilities \cite{Zol2,Cip,Brass1} - the
interplay of the extraordinary and ordinary components of the
irradiating light beam. This effect works only with monochromatic
light beams; the analysis of white light sources is still an open
problem.

\section{Analysis of light scattering}

The description presented in the previous section showed that the
uniform planar orientation must become unstable in the reverse
geometry for $\alpha > 0$. Arguments were given to explain that
why the distortion becomes static near $\alpha = 90^{\circ}$ and
dynamic for $0<\alpha < 90^\circ$. The above considerations,
however, cannot elucidate the underlying nature of the
instabilities, especially not that of the dynamic one.  This task
is beyond the scope of the current paper; we present nevertheless
some empirical results in connection with light scattering. These
findings, although they are preliminary, give some indications of
the class the observed dynamic behavior belongs to.

The autocorrelation function of the scattered light intensity did
not exhibit any structure, in particular it did not show
oscillatory character. This fact seems to be in contrast with the
occurrence of quasi-periodic oscillations that have been reported
in \cite{Jan} on the basis of observations in a polarizing
microscope. Considering the results shown in Fig.3 of Ref. [12],
the oscillations at the given light intensity could be expected to
occur around 1.5Hz. However, in \cite{Jan} a much larger spot was
illuminated than in the laser experiments ($mm$-s versus $100
m\mu$-s). In the former case - after some time - a quasi-periodic
structure was organized, which seems to be absent in the latter
one. We note that the same light-scattering experiment was
repeated with a much higher sampling rate (100 Hz); the result was
similar as with the low sampling rate.

We investigated also the probability density function (PDF) of the
scattered light. As it can be seen from Fig. 1, in a planar cell
the scattered light occasionally increased sharply. These spikes
could be followed by eye also in the scattering picture; they
corresponded to flashes of randomly oriented lines, as mentioned
in Sec.\ II. Such intermittent behavior is well-known to occur in a
great number of disparate systems e.g., in turbulent flow
\cite{Cad,Bram2,Pin}, in Danube water level fluctuations
\cite{Jani,Bram3}, in the simulations of the 2D X-Y model at
criticality \cite{Bram,Bram2,Aum}, or in electroconvection of
liquid crystals \cite{TTK}. In all these systems the probability
density function of the measured quantities follows the
generalized Fisher-Tippet-Gumbel (gFTG) distribution \cite{Bram}
which has a form of
\begin{equation}\label{gFTG}
    \pi (S)\sigma_S=K\exp [b(x-c)-e^{b(x-c)}]^a
\end{equation}
where $K = 2.14$, $b = 0.938$, $c = 0.374$, $a = \pi/2$, and
\begin{equation}\label{x}
    x=(S-<S>)/\sigma_S
\end{equation}
In our case: $\pi(S)$ - PDF of the fluctuations of the light
intensity scattered into the observed area, $S$; $<S>$ - the mean
value of the scattered light intensity; $\sigma_S$ - the standard
deviation of the $S$ fluctuations. Note that gFTG distribution
presented by Eq. (\ref{gFTG}) has a universal form, without any
fit parameter. The gFTG distribution is substantially skewed with
one tail described by an exponential decay which is considered to
be due to fluctuations having a length scale comparable to the
system size \cite{Bram2,Pin,Por}.

In Fig. 3 we present the normalized PDFs deduced from our
measurements with $\alpha  =45^\circ$ and $\alpha  =90^\circ$ of
the incoming light. The solid lines correspond to the Gaussian and
gFTG distributions (without a fit parameter) as denoted in the
legend. Clearly, in case of $\alpha  =90^\circ$ the central limit
theorem holds. In this geometry the source of the fluctuations is
mainly the laser and the dark current of the photomultiplier; the
intensity fluctuations have Gaussian distribution suggesting that
they arise from many uncorrelated contributions. On the contrary,
for $\alpha =45^\circ$ the PDF of the fluctuations follow the gFTG
distribution. This is an indication of an essential difference
between the lights scattered from the static and dynamic domains.

The fluctuations of scattered light have also been analyzed for a
wide range (of almost three orders of magnitude) of input powers
$P$. Fig. 4 shows the temporal dependence of the scattered light
intensity in the range of input powers from $P=12\mu$W to
$P=4.2$mW  for $\alpha=45^\circ$ of the incoming light.
Qualitatively, the temporal dependencies seem self-similar. As a
quantitative measure of the self-similarity we have plotted the
normalized PDFs of these fluctuations in Fig. 5. As one can see
the fluctuations follow satisfactorily the gFTG distribution over
the whole range of the input power investigated.

In  Sec. III it was pointed out that the dynamic instability is
related to the continuous  generation/annihilation of disclination
loops. We assert that the spikes observed in the case of planar
cells arise from disclination lines, passing through the area
illuminated by the exciting light beam. In the restricted
irradiated area spikes are probably due to individual disclination
loops. This circumstance is deduced from the observation by eye,
where the scattered lines are seen individually.

In a recent paper it has been argued that the asymmetry of the
energy distribution is driven by a finite energy flux crossing the
system from injection to dissipative scales \cite{Ber}. In the
same time, as it has been pointed out, a comparison with a real
experimental system (such as the turbulent flow) is not directly
feasible, due to a number of simplifications in the
phenomenological approach \cite{Ber}. Despite of that, some
comments on the experiments can be made in this regard. In our
case the disturbed area in the sample for a focused beam is larger
than the laser spot due to long-range elastic interactions in the
nematic (just as in the case of static optical Freedericksz
transition with finite beam size, where the director distortion
has an essentially longer range than the laser diameter \cite
{Zel,Csill}). Therefore, we assert that disclination loops are
nucleated in a larger area then the scattered one. However, if one
suppose that dissipative scales are connected to the deformation
zone in the director field around the disclination lines only,
then the idea of Bertin and Holdsworth \cite{ Ber} may be relevant
to our system. Future investigations of the PDF dependence on the
laser spot size may help to clarify this question.

As mentioned in Section 2, for hybrid cells the fluctuations are
less pronounced, spikes are basically absent. The PDFs of the
fluctuations for this case are presented in Fig. 6. The data for
all polarization direction collapse into one "intermediate" PDF
(in spite of the intensity differences shown in Fig. 2) which
deviates both from the Gaussian and from the gFTG distributions,
and represents a midway distribution between them. In the case of
hybrid cells the undamped rotation of the director can take place
in a much smoother way than in the planar case; disclination loops
are only generated in the vicinity of the border of illuminated
area. This fact could explain the scarcity of spikes for the
hybrid cell. The situation is analogous  to the distribution of
injected power fluctuations in electroconvection [21]. In those
measurements for the unconfined geometry, a departure from the
normal distribution towards the gFTG one has been observed above
the anisotropic to isotropic turbulence, when the density of the
disclination loops is abruptly increased.

In conclusion, the observed light scattering from chaotic director
movement is analogous to observations of turbulence in many
different systems. Further work is needed to elucidate the
similarities and differences in the details.

\section{Conclusions}

In this paper the previous studies of light-induced instabilities
in cells with a photosensitive substrate were extended to the case
of laser illumination. We found that using a monochromatic light
source dynamic instability can be generated in planar cells by
choosing the polarization angle of the exciting beam in a proper
range. In hybrid cells we observed dynamic instability at
arbitrary angle of polarization, although the director turbulence
was less pronounced in this case than in the former one. These
observations were confirmed by light scattering studies as well. A
simple model was presented, which qualitatively explains the onset
of the instabilities.

It is remarkable that a seemingly trivial extension of the
standard photoalignment  experiments, namely reversing the
illumination direction of the exciting light, leads to the highly
nontrivial phenomena described in the paper. As we pointed out,
the reason for pattern formation is that in the reverse geometry
the photoalignment and the director reorientation processes are
coupled together and their mutual interaction gives rise to the
observed behavior of  liquid crystal cells.

\section*{Acknowledgments}
The work was supported by the Hungarian Research Fund OTKA K81250.

\newpage

\begin{figure}
\begin{center}
\includegraphics[width=30pc]{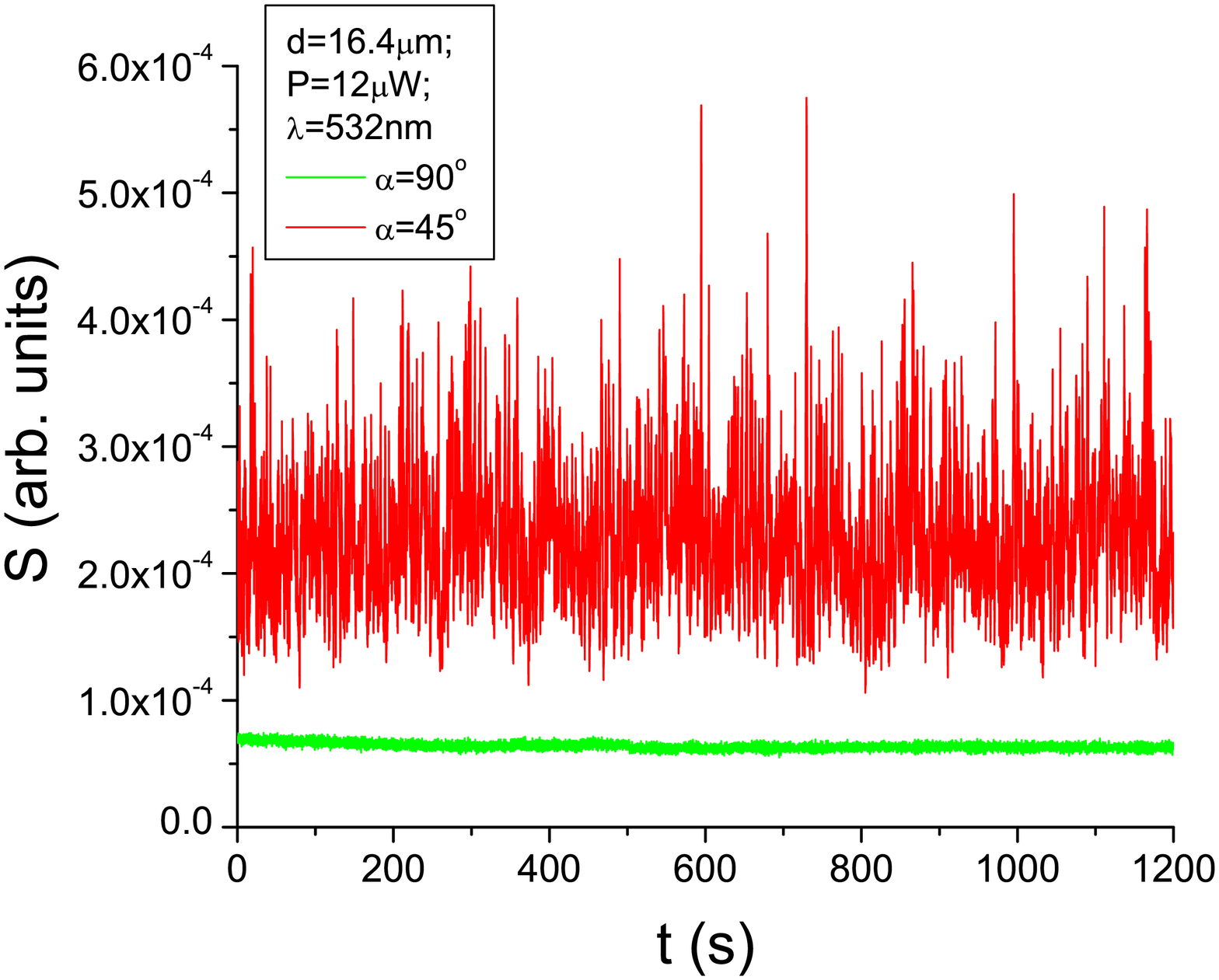}
\caption{(Color online) Temporal dependence of the scattered light
intensity for the polarizations $\alpha=90^\circ$ and
$\alpha=45^\circ$ of the incoming light in the reverse geometry,
measured in a planar cell.}
\label{Figure1}
\end{center}
\end{figure}

\begin{figure}
\begin{center}
\includegraphics[width=30pc]{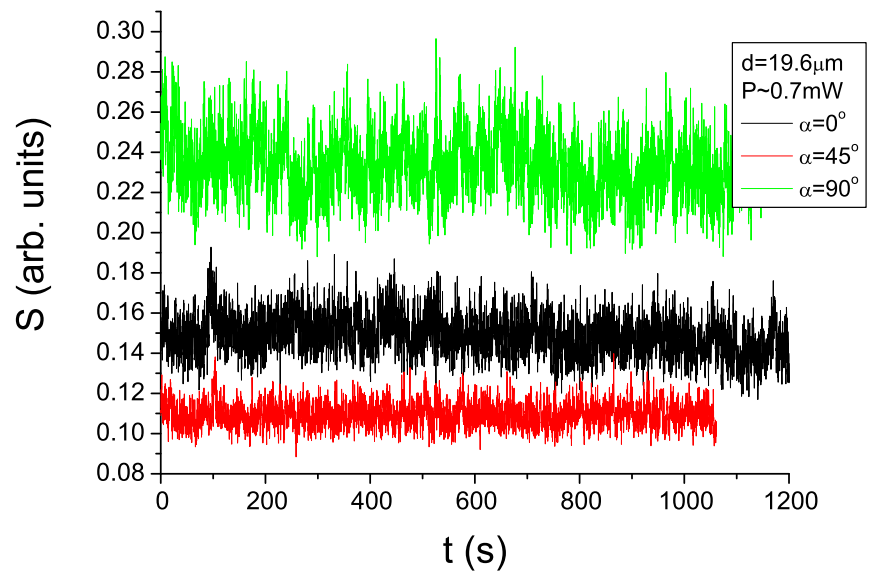}
\caption{(Color online) Temporal dependence of the scattered light
intensity for different  polarizations at the power of about 0.7mW
of the incoming light in the reverse geometry measured in a hybrid
cell.}
\label{Figure2}
\end{center}
\end{figure}

\begin{figure}
\begin{center}
\includegraphics[width=30pc]{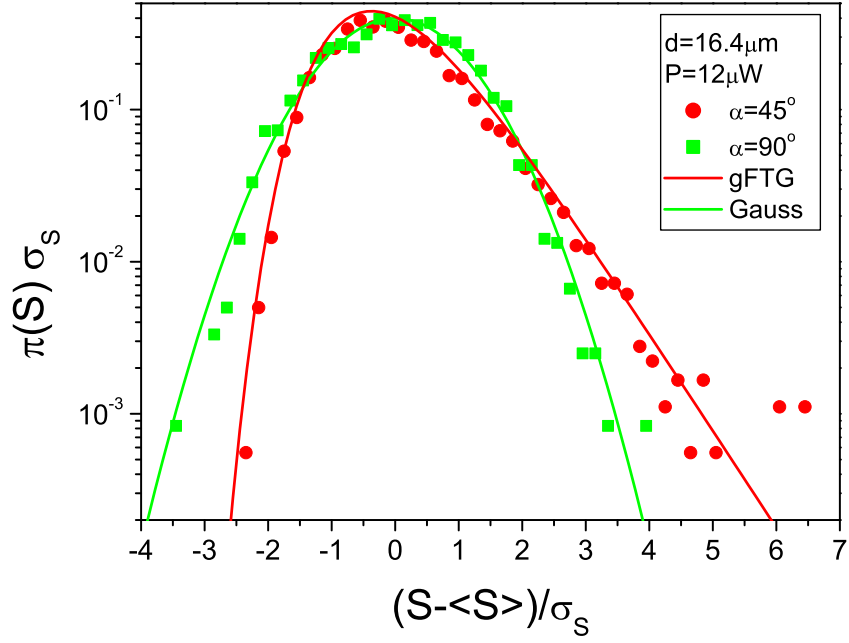}
\caption{(Color online) Normalized probability density function of
the scattered intensity fluctuations detected in a planar cell for
the polarizations $\alpha=90^\circ$ and  $\alpha=45^\circ$ of the
incoming light in the reverse geometry. Solid lines represent
Gaussian and gFTG distributions without fit parameters as denoted
in the legend.}
\label{Figure3}
\end{center}
\end{figure}

\begin{figure}
\begin{center}
\includegraphics[width=30pc]{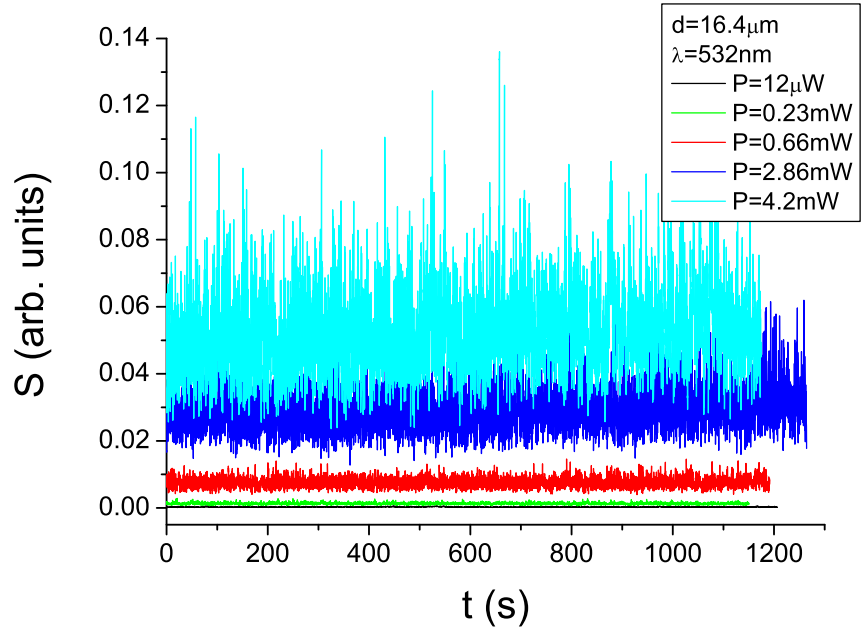}
\caption{(Color online) Temporal dependence of the intensity
fluctuations measured in a planar cell for the polarization
$\alpha=45^\circ$ at different powers of the incoming light in the
reverse geometry.}
\label{Figure4}
\end{center}
\end{figure}

\begin{figure}
\begin{center}
\includegraphics[width=30pc]{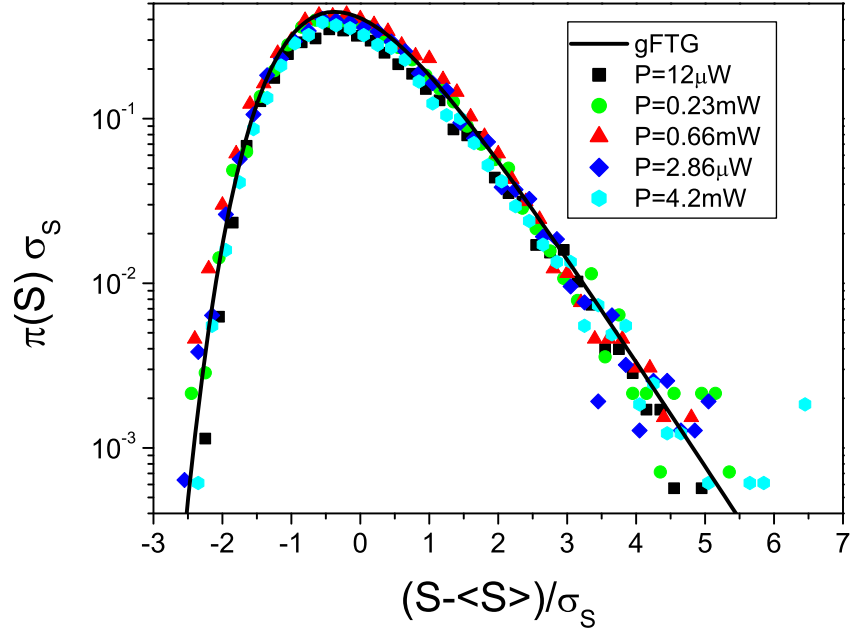}
\caption{(Color online) Normalized probability density function of
the intensity fluctuations detected in a planar cell for the
polarization $\alpha=45^o$ at different powers of the incoming
light (as denoted in the legend) in the reverse geometry. The
solid line represents the gFTG distribution without any fit
parameter.}
\label{Figure5}
\end{center}
\end{figure}

\begin{figure}
\begin{center}
\includegraphics[width=30pc]{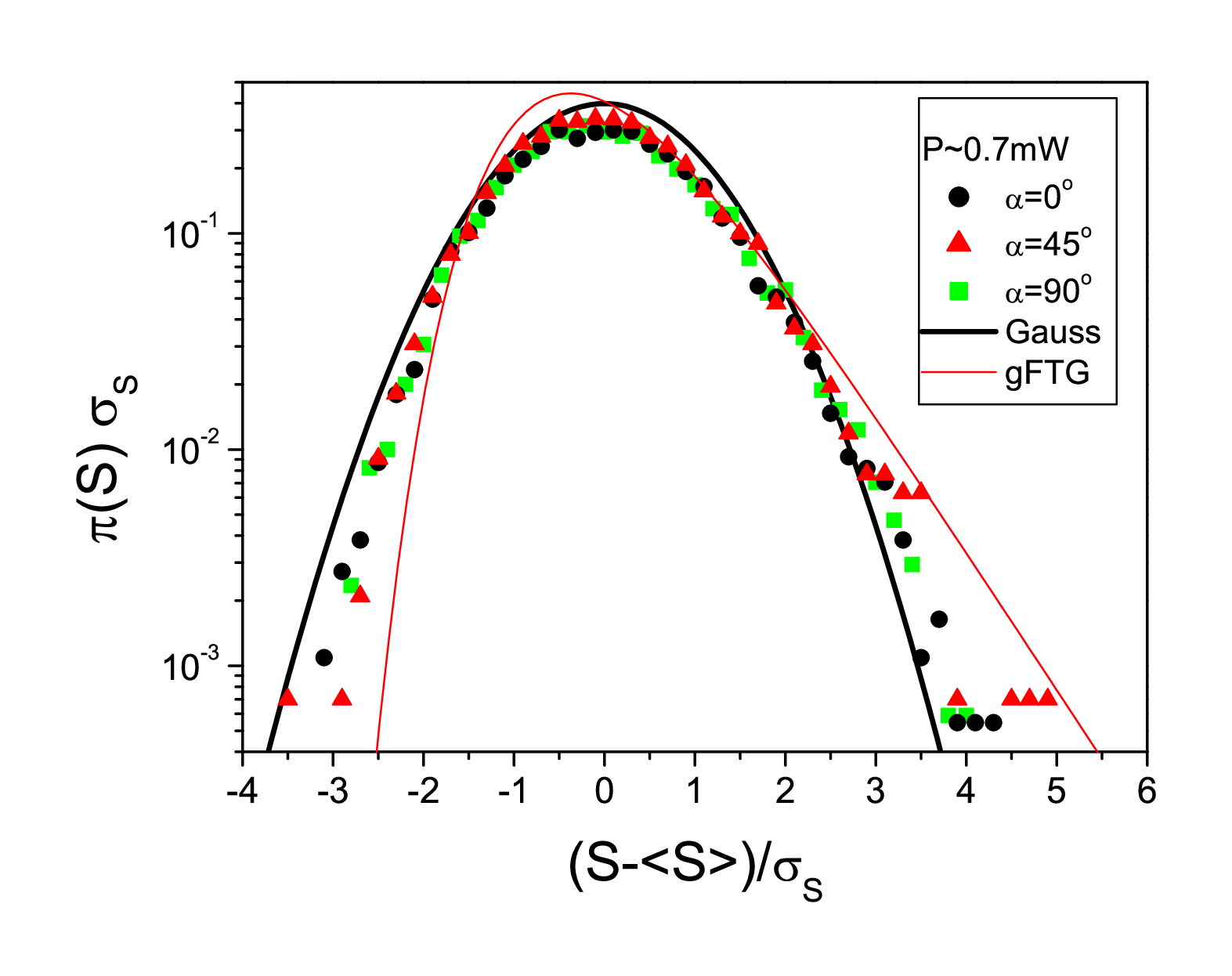}
\caption{(Color online) Normalized probability density function of
the intensity fluctuations detected in a hybrid cell for the power
P$\sim$0.7mW of the incoming light at different polarization
directions (as denoted in the legend) in the reverse geometry. The
solid lines represent the Gaussian and the gFTG distribution
without any fit parameter.}
\label{Figure6}
\end{center}
\end{figure}


\begin{thebibliography}{00}
\bibitem{footnote1} This photoalignment process normally worked for
about a month, afterwards it could not be observed. A possible
reason is that the azo layer was gradually dissolved in the liquid
crystal layer.
\end{thebibliography}
\end{document}